\documentclass[sigconf]{acmart}

\usepackage{url}
\usepackage{hyperref}
\usepackage{graphicx}
\usepackage{booktabs}
\usepackage[T1]{fontenc}
\usepackage{tikz}
\usepackage{svg}
\usepackage{pdfcomment}
\usetikzlibrary{positioning, shapes.geometric, fit, arrows.meta, shadows}

\AtBeginDocument{%
  }

\setcopyright{acmlicensed}
\copyrightyear{2018}
\acmYear{2018}
\acmDOI{XXXXXXX.XXXXXXX}
\acmISBN{978-1-4503-XXXX-X/2018/06}

\newcommand{\approachname}{Code4Me V2}

\begin{document}

\title{\approachname : a Research-oriented Code-completion Platform}

\author{Roham Koohestani\textsuperscript{*}}
\affiliation{%
  \institution{Delft University of Technology}
  \city{Delft}
  \country{the Netherlands}}
\email{rkoohestani@tudelft.nl}

\author{Parham Bateni\textsuperscript{*}}
\affiliation{%
  \institution{Delft University of Technology}
  \city{Delft}
  \country{the Netherlands}}
\email{pbateni@tudelft.nl}

\author{Aydin Ebrahimi\textsuperscript{*}}
\affiliation{%
  \institution{Delft University of Technology}
  \city{Delft}
  \country{the Netherlands}}
\email{aydinebrahimi@tudelft.nl}

\author{Behdad Etezadi\textsuperscript{*}}
\affiliation{%
  \institution{Delft University of Technology}
  \city{Delft}
  \country{the Netherlands}}
\email{betezadi@tudelft.nl}

\author{Kiarash Karimi\textsuperscript{*}}
\affiliation{%
  \institution{Delft University of Technology}
  \city{Delft}
  \country{the Netherlands}}
\email{makarimi@tudelft.nl}

\author{Maliheh Izadi}
\affiliation{%
  \institution{Delft University of Technology}
  \city{Delft}
  \country{the Netherlands}}
\email{m.izadi@tudelft.nl}

\renewcommand{\shortauthors}{Koohestani et al.}

\thanks{\textsuperscript{*}Authors contributed equally.}

\begin{abstract}
The adoption of AI-powered code completion tools in software development has increased substantially, yet the user interaction data produced by these systems remain proprietary within large corporations. This creates a barrier for the academic community, as researchers must often develop dedicated platforms to conduct studies on human–AI interaction, making reproducible research and large-scale data analysis impractical. 
In this work, we introduce \approachname, a research-oriented, open-source code completion plugin for JetBrains IDEs, as a solution to this limitation. \approachname ~ is designed using a client-server architecture and features inline code completion and a context-aware chat assistant. Its core contribution is a modular and transparent data collection framework that gives researchers fine-grained control over telemetry and context gathering. \approachname ~ achieves industry-comparable performance in terms of code completion, with an average latency of 200ms. We assess our tool through a combination of an expert evaluation and a user study with eight participants. Feedback from both researchers and daily users highlights its informativeness and usefulness. We invite the community to adopt and contribute to this tool. More information about the tool can be found on \url{https://app.code4me.me}.

\end{abstract}

\ccsdesc[500]{Human-centered computing~Human computer interaction (HCI)}
\ccsdesc[500]{Software and its engineering}

\keywords{AI-assisted Programming, Research Tooling, Open Science, AI4SE}

\maketitle
\section{Introduction}
\label{sec:introduction}

With the increased use of AI in software engineering, developer workflows are undergoing a fundamental transformation. Industry surveys report widespread adoption of AI coding assistants~\cite{StackOverflowSurvey2025}, with empirical studies demonstrating significant productivity gains; for instance, GitHub Copilot users have been observed to complete tasks up to 55\% faster~\cite{Peng2023Impact}.
Similarly, JetBrains reports that users save up to 8 hours per week with its AI Assistant~\cite{JetBrainsAI2024Survey}, and an Amazon study found that developers using CodeWhisperer completed tasks 57\% faster than those who did not~\cite{AWSCodeWhisperer2023}.

While the benefits of these tools are evident~\cite{Copilotbenefit}, the study of interaction modalities and the wide-scale analysis of developer behavioral data remain limited. This is because the most powerful systems are commercial and closed-source, which presents three critical challenges for academic research: a lack of transparency into the models' decision-making processes, an inability to control experimental conditions (e.g., model versions or data sources), and no access to the rich telemetry data required for fine-grained analysis. Existing academic studies have often required the creation of bespoke, single-purpose tools, consequently restricting the ability to reuse the infrastructure~\cite{Izadi2024Code4Me}~\cite{mozannar2024readinglinesmodelinguser}.

To address this gap, we propose \approachname, an extensible and research-oriented framework for empirical analysis of AI-assisted software development. As an open-source plugin for JetBrains IDEs, \approachname ~ ~ provides core functionalities, such as inline code completion and an interactive chat assistant. Its primary purpose, however, is to serve as a platform for experimentation. The architecture is explicitly modular, and it separates user-facing components from a configurable data collection and model inference backend.

We expect this tool to serve as a useful infrastructure component for the AI4SE community. By offering a shared, transparent, and extensible platform, \approachname ~ enables researchers to focus on experimental design and data analysis rather than on building and maintaining their own data collection systems. In the subsequent sections, we analyze the domain in \autoref{sec:analysis}, describe the design of \approachname ~ in \autoref{sec:approach}, present a preliminary analysis in \autoref{sec:preliminary-analysis}, and discuss limitations and future work in \autoref{sec:limitations}. All relevant links for \approachname ~ can be found at \url{https://app.code4me.me}.

\section{Domain Analysis}
\label{sec:analysis}
The current ecosystem of AI coding assistants is dominated by powerful commercial tools, which, despite their utility, present significant limitations for research. Platforms like \textbf{GitHub Copilot}~\cite{githubcopilot}, \textbf{Cursor}~\cite{Cursor}, and \textbf{JetBrains AI}~\cite{jetbrainsai} offer robust, integrated user experiences. However, their proprietary nature makes them effectively "black boxes." Researchers cannot inspect the suggestion logic, control for model updates that could act as confounding variables, or access raw interaction data.

Open-source alternatives, such as \textbf{Cody}~\cite{Cody} and \textbf{Code4Me}~\cite{Izadi2024Code4Me}, offer more transparency but were not designed with research extensibility as a primary goal. Their architectures often lack modularity, which complicates systematic studies of developer–AI interaction. For example, adapting these systems to support different experimental configurations or data collection strategies typically requires substantial modification of the core codebase.

To better understand these challenges, we analyzed seven widely used coding assistants and their documentation.~\footnote{The platforms analyzed include: 
\href{https://github.com/features/copilot}{GitHub Copilot}, 
\href{https://www.cursor.com}{Cursor}, 
\href{https://cody.dev/}{Cody}, 
\href{https://devin.ai/}{Devin}, 
\href{https://replit.com/site/ghostwriter}{Replit Ghostwriter}, 
{Code4Me}~\cite{Izadi2024Code4Me}, 
and \href{https://www.jetbrains.com/ai/}{JetBrains AI}.} 
Finding that both commercial and open-source options fall short of research needs, we designed \approachname ~ for the academic community, which is transparent, with an open-source platform from plugin to backend; controllable, giving researchers fine-grained control over data collection through a modular system; and extensible, allowing new components such as telemetry modules or context providers to be added with minimal effort.

\section{\approachname}
\label{sec:approach}
\begin{figure}[tb]
    \centering
    \includegraphics[width=\linewidth]{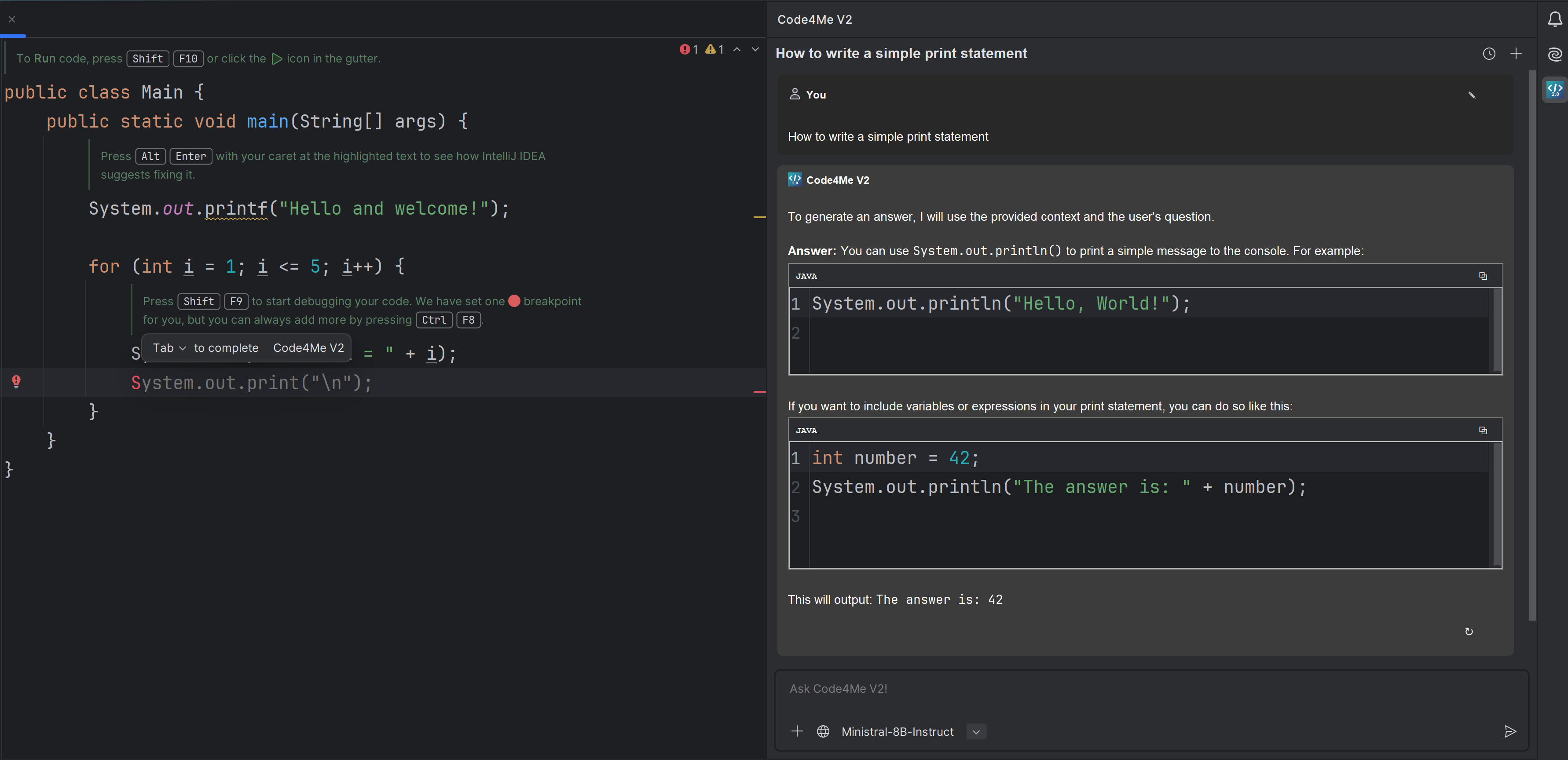}
    \caption{Plugin screenshot (inside IntelliJ IDEA)}
    \label{fig:demo-plugin}
\end{figure}

\begin{figure}[tb]
    \centering
    \includegraphics[width=\linewidth]{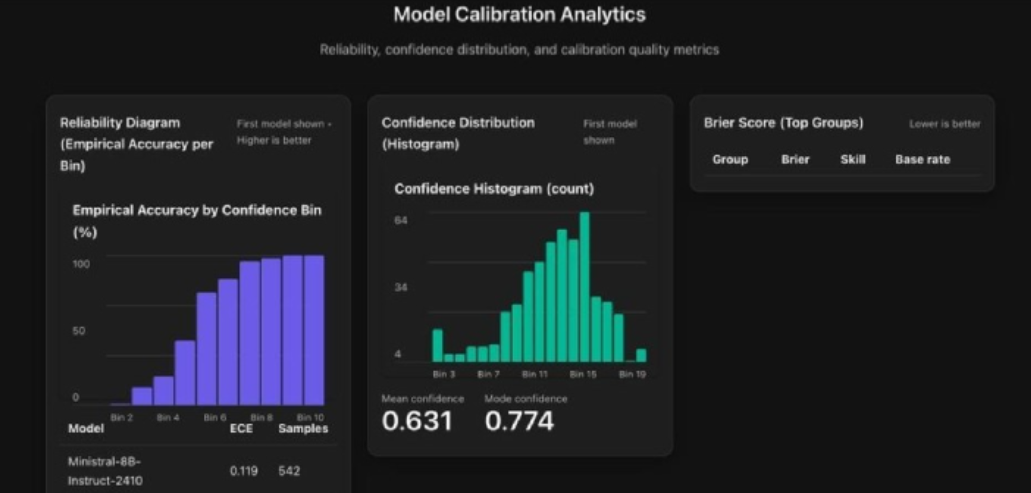}
    \caption{Model calibration pane in the analytics dashboard}
    \label{fig:demo-analytics-dash}
\end{figure}

\approachname ~ is a flexible, extensible research platform built on modularity and separation of concerns. This section outlines its architecture and design, with \autoref{fig:demo-plugin} and \autoref{fig:demo-analytics-dash} showing the plugin and analytics dashboard.

\begin{figure*}[tb]
    \centering
    \includegraphics[width=0.80\linewidth]{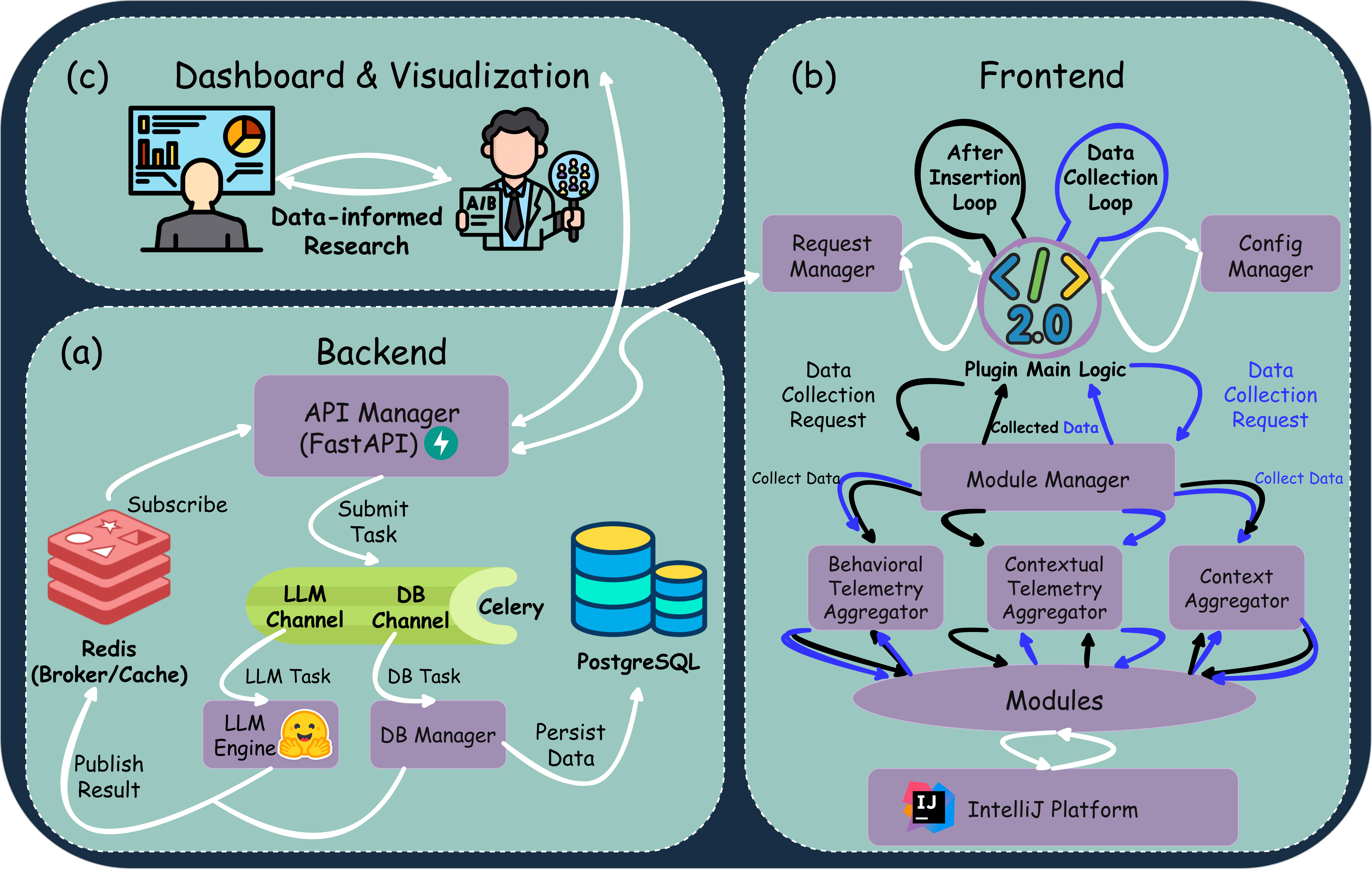}
    \caption{Graphical depiction of the high-level architecture of ~\approachname. (a) represents the backend processes of the server. (b) gives an architectural view of the implemented IntelliJ Plugin for ~\approachname, and (c) represents the admin dashboard.}
    \label{fig:architecture}
\end{figure*}

\subsection{General Architecture}
\label{sec:approach:architecture}
The system is designed using a client-server architecture to enforce a strict separation of concerns. This allows for the different components to be modified separately from each other, and allows for further extension down the line (e.g., by adding a VSCode Client, or adding a new model). In terms of the current implementation, \textbf{The Client} is a lightweight JetBrains IDE plugin. Its sole responsibilities are rendering the user interface (ghost text and chat), collecting user-configurable context and telemetry from the editor, and dispatching requests to the server. \textbf{The Server} handles all heavy lifting: user authentication, persistent data storage in a relational database, and the computationally intensive AI model inference pipeline.
The decision for using the client-server architecture comes into play here as well. It minimizes the performance overhead on the developer's IDE, as all significant computation is offloaded. In addition to these main components, we have developed the \textbf{Analysis Platform}, which serves as the user interface for the researchers. This component is in direct communication with the backend and helps the researcher analyze the collected data, gain insights into usage patterns, and set up user studies.
In the sub-sections to follow, we will go more in-depth into each of these components.

\subsection{Backend}
\label{sec:approach:backend}
The backend is a Python application built with modern, scalable technologies. The API is served by \textbf{FastAPI}~\cite{fastapi2024}, chosen for its asynchronous capabilities, which allow it to handle many concurrent requests efficiently. A critical design choice was to decouple long-running tasks, such as LLM inference and persistence, from each other and from the main API request-response cycle. We achieve this using \textbf{Celery}~\cite{celery}, a distributed task queue, with \textbf{Redis}~\cite{redis2024} as a message broker. When a completion request arrives, the API immediately places it on the task queue for both the inference engine and the database manager. A Celery worker picks up the task, performs the model inference, and publishes the result. This asynchronous design ensures the user's IDE never hangs while waiting for irrelevant processes to the actual completion to finish. The infrastructure provides support for both WebSocket and HTTP-based requests, which in turn gives the front-end developer a lot of flexibility in how they want to implement features.

All data, including user profiles, telemetry, and suggestion metadata, is stored in a \textbf{PostgreSQL}~\cite{postgresql2024} database, with \textbf{SQLAlchemy}~\cite{sqlalchemy2024} as the ORM to ensure data integrity and adherence to ACID principles. A graphical depiction of the processes and responsibilities of the backend components has been presented in part (a) of \autoref{fig:architecture}.

\subsection{Frontend}
\label{sec:approach:frontend}
The frontend architecture is central to \approachname's effectiveness as both a user-focused application and a research instrument. It is centered around a highly modular system. In the design, the core unit is defined by a Module. A module can be responsible for collecting data for the completion, running some update process, or running an arbitrary post-acceptance process. Modules follow a hierarchical system. To gain a better idea, the \textbf{Module Manager} is the central coordinator that receives data collection requests from the plugin's main logic. \textbf{Aggregators} are modules that directly receive dispatched messages from the manager. These are responsible for a specific class of data. Each aggregator invokes one or more individual, self-contained \textbf{Modules}. For example, the \texttt{Behavioral Telemetry Aggregator} might call the \texttt{Typing Speed} and \texttt{Time Since Last Completion} modules. Keep in mind that Aggregators themselves are modules, so this \textit{self-contained module} can also be an aggregator. This composable structure allows researchers and developers to have highly complex interaction systems. 

This design makes the system exceptionally extensible. To add a new piece of telemetry (e.g., tracking copy-paste events), a researcher only needs to implement a simple module interface and register it in the configuration. No other part of the system needs to be modified; this drastically lowers the engineering effort required to design new experiments. In line with the guiding principles of IntelliJ Action systems \footnote{\href{https://plugins.jetbrains.com/docs/intellij/action-system.html\#principal-implementation-overrides}{IntelliJ Action System principles}}, actions initiated by the plugin occur exclusively when a user submits a completion request, whether inline or through the chat module. We classify module actions into two categories: the data collection loop and the after-insertion loop (these actually also correspond to two separate methods in the Module interface). These represent stages within the request lifecycle where modules can integrate and execute necessary actions. The complete lifecycle of this process is detailed in part b of \autoref{fig:architecture}.

\subsection{Analytics Platform}
\label{sec:approach:analysis-platform}
The analytics subsystem transforms telemetry collected from various stack layers into research-quality metrics and summaries. The event streams, which include meta\-queries, model generations, and contextual and behavioral telemetry, are stored in PostgreSQL and examined through SQL window functions and time bucketing methods. The public API provides endpoints that calculate time‑series aggregates, acceptance statistics with confidence intervals, and latency percentiles. This supports both descriptive and comparative analysis of model performance in realistic scenarios. Role-based access control is uniformly enforced; non-admin users have access to their analytics, whereas admins can request cross‑user comparisons and handle experiments and configurations. Model-specific endpoints allow for side‑by‑side comparisons of acceptance, latency, and confidence distributions, along with calibration analyses. These metrics allow for studies into confidence validity and the correlation between predicted confidence and actual acceptance. Additionally, the system supports A/B testing and configuration management to enable controlled platform experiments. Only admins can access study endpoints, which are responsible for lifecycle management and random server-side configuration assignments.

\section{Preliminary Evaluation}
\label{sec:preliminary-analysis}
The platform has been evaluated for performance, stability, and its suitability for research. The system's architecture was designed to minimize client-side overhead by offloading all model inference to the server. 
To assess the server-side latency, we create a pipeline using a recognized Fill-in-the-Middle task dataset~\footnote{\href{https://huggingface.co/datasets/bigcode/santacoder-fim-task}{Santacoder FIM task.}} and continuously send requests to the server to measure the average delay. In the same manner, for the chat functionality, we use a well-known program synthesis dataset~\footnote{\href{https://huggingface.co/datasets/google-research-datasets/mbpp}{We use the MBPP dataset.}} to instruct the model to clarify and execute the task, mimicking a scenario where a user is inquiring from within an IDE. For code completion, we use \textit{deepseek-coder-1.3b-base}, and for the chat model, we use \textit{Ministral-8B-Instruct}. End-to-end latency for code completion requests, from the trigger event in the IDE to the rendering of ghost text, is consistently 186.31 ms (± 139.50), with an average of 18.66 tokens, which is well within the threshold for a non-disruptive user experience. For chat completions, the latency is 8369.78 ms (± 840.48) with an average of 277.08 tokens.

To evaluate the plugin's usability and research extensibility, we conducted a two-phase user evaluation. First, a formative user study was conducted with four expert researchers from the AI for Software Engineering (AI4SE) faculty. Participants were deliberately chosen based on their familiarity with AI4SE and their frequency of using AI coding assistants. They were then asked to complete a basic multi-file programming task, which involved creating a `Shape` interface and `Circle` and `Rectangle` implementing classes. During this task, they were expected to interact with both the inline completion and chat features.

Feedback was gathered through a semi-structured interview format and a questionnaire. The questionnaire focused on three key areas: \textbf{General Usability} (e.g., relevance and timeliness of suggestions), \textbf{Setup \& Configuration} (e.g., clarity of data collection settings), and \textbf{Extensibility for Research} (e.g., modifying the plugin for experiments). The qualitative feedback from this initial study strongly validated the platform's core premise: participants unanimously praised the modularity and extensibility, and expressed confidence in their ability to adapt the tool for their own research workflows. The study also identified areas for improvement, primarily regarding the intuitiveness of the module management UI and inconsistencies in the timing of automatic suggestions.

Based on the results of this formative study, we made several iterations on the initial design. We then conducted a secondary evaluation with four daily users using the same protocol. Our findings revealed that the individuals did not echo the issues previously identified by experts. Furthermore, fewer participants mentioned timing issues with the suggestions, experiencing fewer challenges with frequency and response time. Nonetheless, one user noted that the suggestions could benefit from a slightly increased speed. Lastly, a participant requested the inclusion of an \emph{Agent} feature in the plugin, highlighting (1) a rapidly evolving domain and (2) the necessity for quick adaptation from academia, the later of which is now facilitated by \approachname.

\section{Limitations and Future Work}
\label{sec:limitations}
Our preliminary analysis also identified several limitations. As noted in our user study, the user interface for experiment configuration, while powerful, requires a learning curve that could be steep for non-expert users. While improved in the second round of iterations, there is still room for improvement, both in terms of how settings are shown and whether settings should be shown at all. While the platform is performant for research purposes, its completion speed and model quality do not yet match that of well-resourced commercial tools. The current context retrieval, while rather effective in single-file and multi-file settings, still requires attention to be on par with commercial counterparts. A primary objective is the implementation of project-wide context retrieval to enhance the accuracy of code suggestions. Moreover, continuously improving and modifying the current framework is crucial to continue meeting the community's needs (see the request for an agent feature in \autoref{sec:preliminary-analysis}). Therefore, we invite community members to engage with this project to ensure mutual benefit. We will be overseeing the repositories and welcome community contributions.

\section{Conclusion}
\label{sec:conclusion}
\approachname ~is an open-source, modular platform designed to lower the barrier for empirical research in AI-assisted software development. Our preliminary evaluations demonstrate that the platform is both practical for day-to-day use and adaptable for experimental needs. By providing a transparent, controllable, and extensible tool, we enable the academic community to rigorously investigate the complex dynamics of human-AI collaboration in programming. We believe \approachname ~can serve as a foundational piece of infrastructure for the next wave of AI4SE research.

\bibliographystyle{ACM-Reference-Format}
\bibliography{main}

\end{document}